# Waveform Timing Performance of a 5 GS/s Fast Pulse Sampling Module with DRS4[*]

WANG Jin-Hong(王进红)[1,2]   LIU Shu-Bin(刘树彬)[1,2]   AN Qi(安琪)[1,2]

[1]State Key Laboratory of Particle Detection and Electronics, University of Science and Technology of China, Hefei 230026, China

[2]Department of Modern Physics, University of Science and Technology of China, Hefei 230026, China

**Abstract:** We first clarify timing issues of non-uniform sampling intervals regarding a 5 GS/s fast pulse sampling module with DRS4. Calibration strategy is proposed, and as a result, the waveform timing performance is improved to be below 10 ps RMS. We then further evaluate waveform-timing performance of the module by comparing with a 10 GS/s oscilloscope in a setup with plastic scintillators and fast PMTs. Different waveform timing algorithms are employed for analysis, and the module shows comparable timing performance with that of the oscilloscope.

## 1  Introduction

There are several TOF systems in high energy experiments utilizing long plastic scintillator bars [1]-[3]. In these systems, the two ends of each bar are read out via PMTs, and corresponding pulses from PMTs are transmitted to front-end signal processing circuits as leading-edge discriminators for the arrival time of particles. As the progress of modern technology, it is now conceivable to read out the scintillators with fast PMTs for better timing performance, and upgrade on associated readout electronics is also in demand. It is pointed out in [4]-[5] that waveform sampling gives the best timing precision compared with conventional timing techniques as: leading edge discriminators, constant fraction discriminators, and multiple threshold discriminators. Traditionally, one uses analogue-to-digital converters (ADCs) for pulse sampling in physics experiments [6]-[12]. Recent literatures show that waveform sampling with switched-capacitor arrays (SCAs) is also a promising technique in consideration of system densities, power consumption and financial cost [13]-[19]. Up to the present, several Application Specific Integrated Circuits (ASICs) of SCAs for high-energy physics experiments have been developed [18]-[22]. A review of the representative SCAs can be found in [23].

In our previous work [23], we chose DRS4 [22], [24], the fourth version of Domino Ring Sampler (DRS) from Paul Scherrer Institute (PSI), Switzerland, and built a 5 GS/s fast sampling module. The module is proved to be capable of sub-10 ps RMS waveform timing after a series of calibration strategies [23]-[25]. In this work, we first continue our effort to clarify issues regarding non-uniform sampling intervals of the module. Then we evaluate its timing performance in a cosmic ray setup with plastic scintillators and fast PMTs. The timing performance is also compared with that of a Lecroy 10 GS/s oscilloscope [26] in a similar setup for evaluation of possibilities to improve timing performance of TOF systems [27].

[*]This work was supported by the Knowledge Innovation Program of the Chinese Academy of Sciences (KJCX2-YW-N27), and the National Natural Science Foundation of China (No. 11175176).

We arrange this paper as follows. In section 2, we clarify issues regarding uneven sampling intervals of the module. In Section 3, we evaluate the timing performance of the module by putting it in a cosmic ray telescope with plastic scintillator bars and PMTs, and compare the timing performance with that of the 10 GS/s oscilloscope. Discussions are given in Section 4. Finally, in Section 5, we conclude this paper, and summarize what we have achieved.

## 2 Timing Issues Regarding the Module

There are several factors limiting the potential timing performance of the module, as the analog input bandwidth, maximum sampling rate of the module, as well as the performance optimization of DRS4. Generally, a higher bandwidth and sampling rate results in better timing performance [4]-[5]. The analog bandwidth of DRS4 is as high as 950 MHz [24]. However, it will drop dramatically without a proper arrangement of the input driving circuits for the heavy capacitive load at its input. In the module, the achieved bandwidth is around 600 MHz with fully differential amplifiers. Besides, we operate the module at around the highest sampling rate of DRS4: 4.7 GS/s per channel. Fig.1 is a photograph of the module. A detailed description of the module is given in [23].

Once the module is fabricated, its analog bandwidth and the maximum sampling rate are relatively fixed. We improve its timing precision by optimizing the performance of DRS4, as DC offset compensation, and uneven sampling intervals calibration. The DC offset is the variation of residual voltage in each sampling cell of DRS4, and this variation after compensation can be as low as 0.35 mV RMS [23]. The uneven sampling intervals of DRS4 are a bit more complex to calibrate. In [23], we proposed to do this with zero crossing of sine. In the signal processing, the sinusoidal samples are pre-processed with a low-pass filter before applying the zero-crossing algorithm. The sampling intervals obtained show very small variation (~ 5 ps RMS at 4.7 GS/s), and good performance of the module is achieved after uneven sampling interval calibration and noise suppression. However, we find that the distribution of the sampling intervals obtained in [23] cannot reflect the real delay variation of the domino taps in DRS4, in spite of the good performance achieved. We clarify this as follows.

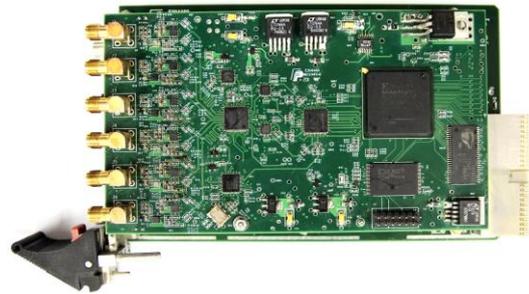

Fig. 1. A photograph of the fast sampling module.

### 2.1 Variation of Sampling Intervals of DRS4

We determine the sampling intervals of DRS4 with zero-crossing of sine waves [23]. The frequency of sine is 100 MHz, and it is sampled at 4.7 GS/s. In [23], a low-pass filter was applied to the sinusoidal samples before performing the zero-crossing algorithm, whereas in this section, the filter is removed and raw sinusoidal samples are used for analysis. We plot several trials of sine waveforms with respect to the sampling cell number in Fig.2. The samples of each trial are arranged in an ascending sequence of the 1024 sampling cells in DRS4 (from cell 1 to 1024). For clarity, we show only part of the waveforms (from cell 800 to cell 850). Theoretically, the sampled sine waves should be smooth everywhere with uniform sampling. However, we observe that there is an up-down alternation of the samples in Fig.2. Moreover, this upward or downward trend at a sampling cell is constant for samples within the same rising or falling edges, e.g., samples at cell 810 show an upward trend in the falling edges, whereas those at cell 830 exhibit a downward tendency in the rising edges.

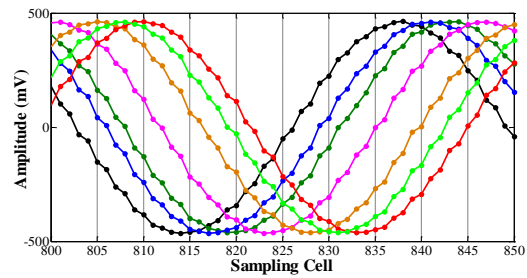

Fig. 2. Up-down alternation of sampled sine points.

For a given sine: $V(t)=V_0 \times \sin(\omega t+\varphi)$, where $\omega$ is the angular frequency and $\varphi$ is its phase. The voltage difference ($\Delta V$: delta V) between two adjacent sampled points across zero is constant in an ideal case (uniform sampling and no voltage distortion): $\Delta V = \omega V_0$. We collect the zero-crossing voltage difference of each sampling cell, and plot the results in Fig.3. Fig.3 (a) shows $\Delta V$ at each sampling cell. The voltage difference per cell is given by an average of hundreds of trials. Fig.3 (b) presents the corresponding distribution of $\Delta V$ for the total 1024 sampling cells. We observe that the voltage difference alternates cell by cell, and they spread into two distributions: one centralizes at about 81 mV, and the other concentrates at about 40 mV. Besides, the variations of the two distributions are both around 6.7 mV RMS. Fig. 3 (c) shows the standard deviation of $\Delta V$ at each sampling cell. The variation of $\Delta V$ at each cell is around 1 mV RMS, which reflects the corresponding zero-crossing voltage difference is quite stable. We can therefore derive the sampling intervals from these voltages.

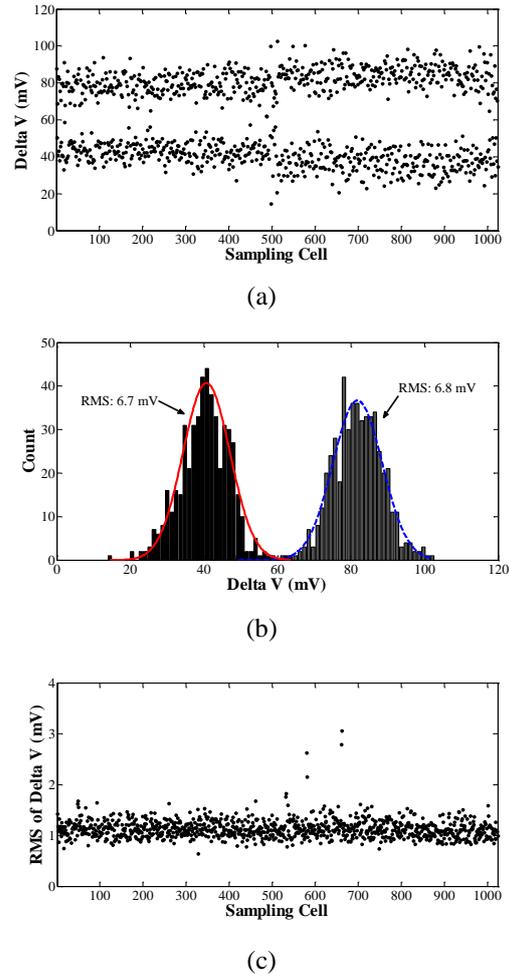

Fig. 3. Voltage differences with zero-crossing method, where (a) illustrates the zero-crossing voltage difference at each sampling cell (cell 1-1024), (b) shows the corresponding distribution, and (c) presents RMS of the zero-crossing voltage difference in (a) for each cell.

At the zero-crossing point of sine, the voltage difference is proportional to the sampling intervals. Therefore from the voltage differences in Fig.3 (a), we can derive the ratio of the 1024 tap delay in DRS4. Taking into account the total sampling intervals equal $1024 \times 1/f_s$ ($f_s$ is the sampling rate) [24], the sampling intervals at each cell can be deduced accordingly. Fig.4 shows the sampling intervals obtained at 4.7 GS/s, where Fig.4 (a) presents the delay of 1024 sampling taps in DRS4, and Fig.4 (b) plots corresponding distribution of the delay. The sampling intervals alternate cell by cell, and the delay spreads into two distributions: one centralizes at 285.8 ps with a RMS of 23.6 ps, and the other converges at about 140.4 ps with 23.5 ps standard deviation. The average delay of the former distribution is about two times that of the latter, and the total variation of the 1024 sampling intervals is about 76 ps RMS. The results are consistent with those in Fig.3.

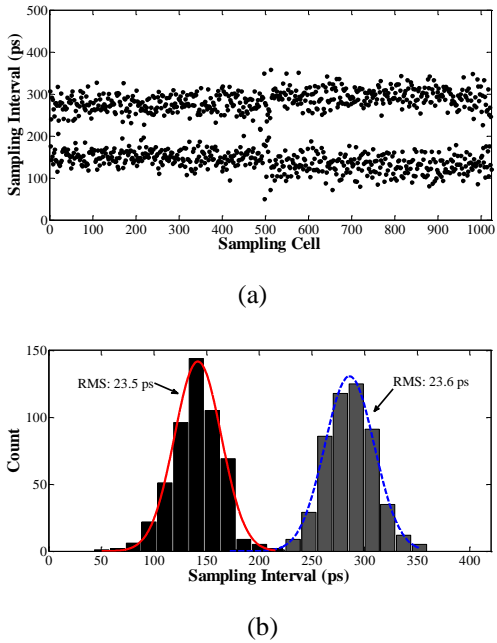

(a)

(b)

Fig. 4. Sampling intervals of DRS4 at 4.7 GS/s, in which (a) shows the sampling intervals at each sampling cell, and (b) presents the corresponding distribution.

## 2.2 Verification of the Sampling Intervals

The sampling intervals derived from prior section show very large variation and spread into two distributions. Thus waveform digitization with the module is subject to non-uniform sampling. If we assume the sine are uniformly sampled and perform spectrum analysis of the raw samples, there will be large distortions at $\pm f_{in} + f_s/2$ ($f_{in}$ is the frequency of sine) [28]. This is verified by the spectrum analysis in Fig.5 (a). There is a large distortion at about 2.247 GHz (marked with a red arrow), which is about $f_s/2 - f_{in} = 4.7\ GHz/2 - 100\ MHz = 2.25\ GHz$. We then interpolate the raw samples with uniform sampling intervals, and plot the corresponding spectrum in Fig.5 (b). We observe the large distortion due to non-uniform sampling disappears. Besides, the signal to noise and distortion ratio (SINAD) is also improved from 31 dB to 41.5 dB. The improvement on spectrum performance reflects that the sampling intervals derived in Fig.4 are a good representation of the actual sampling delay in DRS4. Note there are second and third order distortions in Fig.5. These distortions come from the sine signal generator and the band-pass filters used in our test. The spectrums in Fig.5 are performed by averaging about 200 individual FFTs of 100 MHz sine samples.

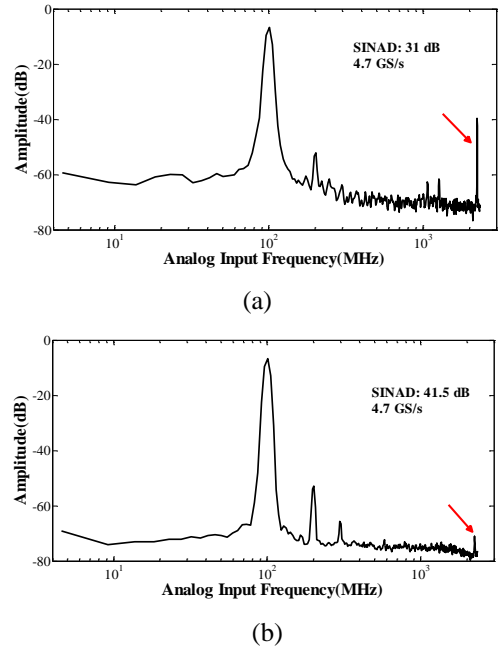

(a)

(b)

Fig. 5. Spectrum analysis of the sine samples, in which (a) shows spectrum of sine before calibration of uneven sampling interval, whereas (b) presents that after calibration.

## 2.3 Timing Performance with the Sampling Intervals

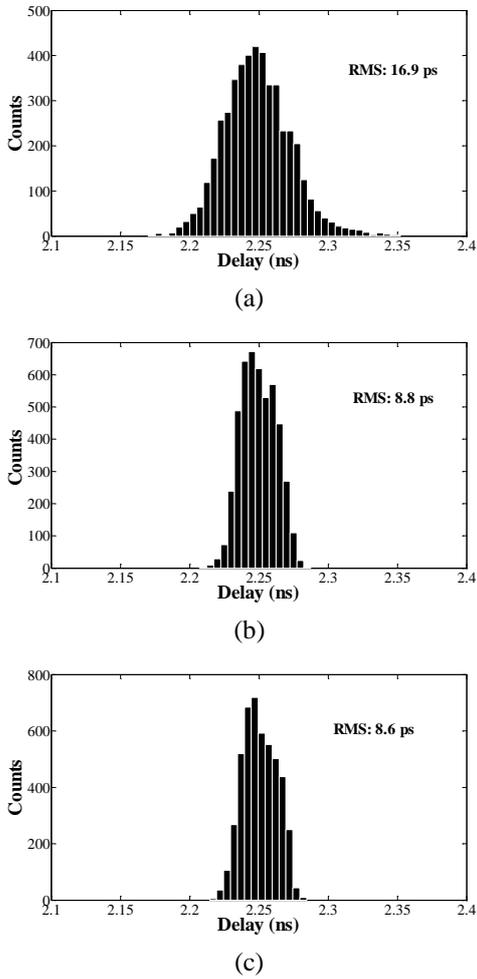

Fig. 6. Distribution of the pule delay derived from waveform sampling, where (a) shows the delay distribution before calibration of the non-uniform sampling intervals, (b) presents that after calibration, and (c) illustrates the time distribution with both low-pass noise suppression and uneven sampling intervals compensation.

We apply the sampling intervals in Fig.4 to the pulse delay test in [23]. In the test, one pulse is split into two with additional delay in one of them. The module samples the pulses, and we perform a 6-order polynomial fitting of the leading edges. The arrival time of a pulse is derived from the crossing time of a digital threshold (200 mV) on the leading edges. Fig.6 shows the time delay, in which Fig.6 (a) presents the time distribution before uneven sampling intervals calibration, Fig.6 (b) depicts that after calibration, and Fig.6 (c) illustrates the time distribution with both calibration of uneven sampling intervals and low-pass suppression as in [23]. The timing performance is improved from 16.8 ps RMS to 8.8 ps RMS after non-uniform sampling interval calibration. There is no significant improvement on timing performance after further processing the calibrated results with low-pass noise suppression (8.6 ps RMS). From this point of view, the low pass noise suppression filter in [23] is no longer essential after aligning the samples with non-uniform sampling intervals derived here.

## 3 Waveform Timing Performance of the Module in a Cosmic Ray Telescope

We built a similar cosmic ray telescope with plastic scintillators and fast PMTs as [27], and put the module in this setup for readout. Different algorithms are employed for waveform timing analysis, and the timing performance of the module is compared with that of the oscilloscope in [27].

### 3.1 Setup of the Experiment

The setup of the experiment is shown in Fig.7. Two identical plastic scintillator bars (EJ200 [29]) of 2360 mm long and 50 mm chick are placed one over the other. The four ends of the scintillators are coupled via four GDB60 PMTs [30] (PMT1-4, 900 ps rise time), and pulses from them are transmitted to the module for digitization. Cosmic rays strike EJ200 from a wide range of solid angles. However, we only choose the portion passing through in the middle for a better characterization of the timing performance [27]. The selection is done with coincidences of pulses from two scintillator-PMT pairs placing in the middle of EJ200 (Scintillator: BC-420 [31]; PMT5, PMT6: XP2020 [32]). Anytime there is a coincidence, the module will be triggered to record the pulses from PMT 1-4. Typical waveforms are shown in Fig.8. The area of each BC-420 scintillator is around 50 mm× 50 mm, and is relatively small with respect to the area of EJ200. Therefore EJ200 can be considered to be bombarded vertically in the middle by the selected cosmic rays.

We extract the arrival time of pulses from PMT 1-4, $t_1$, $t_2$, $t_3$, and $t_4$, from their waveforms, and estimate the timing performance from the standard deviation of the averaging time (t) defined as follows:

t= [($t_1$+$t_2$)-($t_3$+$t_4$)]/4        (1)

The definition in (1) reduces the variation of bombing positions of cosmic rays, and the uncertainty of the referencing time [27].

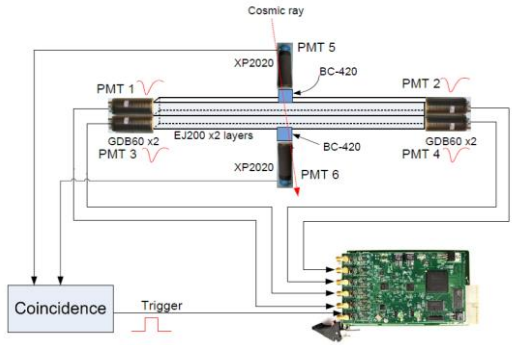

Fig. 7. Experimental setup of the cosmic ray telescope

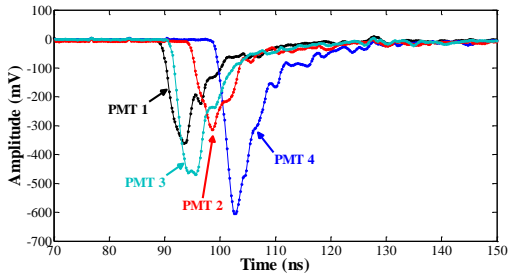

Fig. 8. Typical waveforms from four ends of scintillators (EJ200).

## 3.2 Waveform Timing Algorithms

There are several techniques for time extraction from sampled waveforms, as leading edge discrimination with one or multiple thresholds, digital constant fraction zero-crossing, and pulse shape fitting. A good introduction and comparison of them is given in [4], [33]-[34]. Among these algorithms, some require a constant shape of detector signals, such as $\chi^2$ approach and optimal filtering [34]. These are not suitable for our application, since we sample the waveform without any shaping circuits. Though it is possible to implement pulse shaping with digital signal processing, we are more interested in the information carried by the original waveforms. The algorithms we will use are: digital constant fraction discrimination (d-CFD), cross correlation, and amplitude-weighted sliding window.

d-CFD derives the arrival time from the crossing time at a constant fraction of the pulse amplitude. Cross correlation is a measure of the similarity of two pulses. With the setup in Fig.7, we perform cross correlation of pulses from PMT 1 and PMT 3, PMT 2 and PMT 4, respectively. The time corresponding to the maximum point in the cross correlation waveform represents the time difference, i.e., $t_1$-$t_3$, and $t_2$-$t_4$. Averaging time t is then obtained as (1). In our evaluation, cross correlation of two vectors x and y is calculated from the inverse Fast Fourier Transform of the product: $X^*(e^{jw}) \times Y(e^{jw})$. $X(e^{jw})$ and $Y(e^{jw})$, are the Fast Fourier Transform (FFT) of x and y, respectively. The operator * on the top right of A ($A^*$) computes the complex conjugate of A.

Amplitude-weighted sliding window extracts the arrival time of pulses ($t_d$) from amplitude weighted time in a defined time range (time window: $w_s$). The arrival time is computed as (2).

$$t_d = \frac{\sum_{i=i_0}^{i_0+w_s-1} s_i \times t_d(i)}{\sum_{i=i_0}^{i_0+w_s-1} s_i} \qquad (2)$$

In (2), $i$ is the sample index, starting from $i_0$ and covering a window size of $w_s$. $s_i$ and $t_d(i)$ are the amplitude and time stamp of the $i$-th sample. A detailed introduction of this algorithm can be found in [11].

Table 1. Comparison of waveform timing performance

| ALGORITHMS | TIMING PERFORMANCE (RMS: ps) | | | |
|---|---|---|---|---|
| | THE MODULE | | OSCILLOSCOPE IN [27] | |
| | RAW | CALIBRATED | 10 GS/s | 5 GS/s |
| d-CFD | 52 | 48.5 | 52.3 | 52.1 |
| SLIDING WINDOW | 53.8 | 52.8 | 48.7 | 51.1 |
| CROSS CORRELATION | 57.2 | 55.8 | 61.0 | 59.3 |

### 3.3 Waveform Timing Performance

We evaluate waveform timing performance of the module in the cosmic ray telescope with the three algorithms mentioned in Section 3.2, and compare the results with that of the oscilloscope [27] in Table 1. For consistency in comparison, the raw data of oscilloscope in [27] are reprocessed in an identical way as that of the module, and the obtained timing performance is used for comparison.

In Table 1, the timing performance is for two ends readout of a scintillator bar, therefore the timing variation is $1/\sqrt{2}$ of that in (1). For the category *The Module*, we list waveform timing performance without and with uneven sampling intervals calibration (denoted as *Raw* and *Calibrated* respectively). For the *Oscilloscope*, we also show the timing performance at 5GS/s in addition to that at 10 GS/s. The oscilloscope works at 10 GS/s, and 5 GS/s is achieved by extracting one sample out of every two samples.

For d-CFD, we apply a fourth-order polynomial fitting of samples within 0.05%-30% height in leading edges, and the arrival time is derived from the crossing time of 15% of the pulse height. For amplitude-weighted sliding window, we also choose samples within 0.05%-30% pulse height in leading edges for calculation.

Cross correlation is performed by first interpolating the sampling step to be 20 ps for better precision. Interpolation is done via spline function in Matlab [35].

We observe in Table 1 that the timing performance achieved with the module and the oscilloscope are both around 50 ps RMS. A typical time distribution is shown in Fig.9. There is a slight improvement on timing performance after non-uniform sampling interval calibration for the module. The timing performance for the oscilloscope at 10 GS/s and 5 GS/s are also comparable. The waveform timing precision of the module is proved to be about 10 ps RMS, and it is negligible with respect to the timing variation in Table I (~ 50 ps RMS). Therefore, Table I reflects the potential waveform timing precision of the setup in Fig.7.

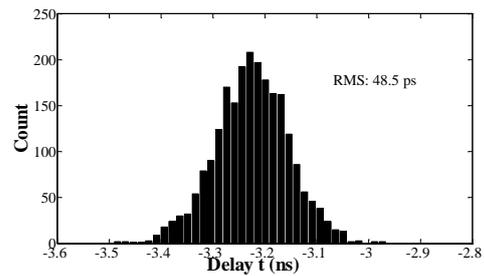

Fig. 9. Typical time distribution of t.

## 4 Discussion
### 4.1 Sampling Intervals Variation of DRS4

The sampling intervals of DRS4 derived in Section 2.1 show much larger variation than those with a pre-signal processing of low-pass filtration in [23]. Good performances are achieved in both cases. However with verification of spectral analysis in Section 2.2, we consider the sampling intervals derived here is a more accurate representation of the sampling tap delay in DRS4. This is consistent with a recent report about DRS4 from Dr. Stefan Ritt [36]. Moreover, with such sampling intervals, no noise suppression strategies are required for the module to achieve sun-10 ps RMS timing precision.

Table 2. Comparison of the module and the oscilloscope

| Parameter | The Module | The Osicilloscope [27] |
|---|---|---|
| Sampling Rate | 4.7 GS/s (max.) | 10 GS/s |
| Bandwidth | 600 MHz | 1 GHz |
| Channels | 6 | 4 |
| Dynamic Range (Vpp/Vrms)[a] | 1 V / 0.35 mV ≈2860 [23] | 10 V /25 mV ≈ 400 [27] |
| Power | ~17.5 mW/channel for DRS4 at 2 GS/s [24] | ---------------- [b] |
| Sampling depth | 1K /channel [24] | 1M/channel [26] |

[a] Vpp is the input voltage range, e.g., for the module in Fig.7, Vpp =1 V. Vrms is the voltage noise for DC input.
[b] No literatures found on the power of ADC used in the oscilloscope.

Compensation of non-uniform sampling intervals of DRS4 is essential for applications with 20 ps RMS timing precision or less. However for those with timing precision of 50 ps RMS or above, such compensation is not essential since no significant timing performance improvement will be obtained. This is reflected in Table 1 for the setup in Fig.7.

## 4.2 Potential Applications in TOF experiments

The timing resolution of current TOF systems has been in the order of 100 ps for several decades [1]-[2], [37]-[38], e.g., the timing resolution is 78 ps in BESIII barrel TOF system [38]. In [27], it is proved that waveform timing is very promising to improve timing resolution in TOF systems. The authors used an oscilloscope in a similar setup as Fig. 7, and a timing resolution of ~50 ps is achieved. In Section 3, we found comparable timing performance can be obtained with the module. We summarize the comparison as Table 2.

From comparison in Table 2, we find the module with DRS4 features such advantages as high channel density (6 of the 8 channels in DRS4 are used in current module), low power consumption, and high dynamic range for input with respect to the oscilloscope. From these points of view, modules with DRS4 can be a good candidate for future TOF upgrade with waveform sampling. There are also drawbacks for DRS4 as limited sampling depth and larger dead time for readout (in the order of one micron second dependent on working mode of DRS4) [24]. However, these will no longer be a problem for DRS5, the fifth version of DRS [39]. Besides, we can also integrate the timing algorithms as amplitude-weighted sliding window and cross correlation on readout electronics. In this way, we are able to alleviate the requirement on data transmission bandwidth by sending the extracted time, instead of the whole waveform.

## 5 Conclusion

We clarified non-uniform sampling intervals of a fast pulse sampling module with DRS4. The sampling intervals were derived by zero-crossing of sine, and verified from the view of spectrum analysis. We then evaluate the performance of the module in a cosmic ray setup with plastic scintillators and fast PMTs. Different algorithms are used for waveform timing analysis, and the timing performance is comparable with respect to a 10 GS/s oscilloscope in a similar setup.

## Acknowledgment

The authors would like to thank S. Li, and Y. Heng from Institute of High Energy Physics, Chinese Academy of Science, for their help regarding this work.